\documentclass[journal]{IEEEtran}

\usepackage{epsfig,amsmath,amssymb,epsf,amsthm,scalefnt,multirow}
\usepackage{xcolor}
\usepackage{float}
\usepackage[caption=false,font=normalsize,labelfont=sf,textfont=sf]{subfig}
\usepackage{textcomp}
\usepackage{stfloats}
\usepackage{cite}
\usepackage{psfrag}
\usepackage{bm}
\usepackage{algorithmic,algorithm}
\usepackage{graphicx}
\usepackage{mleftright}

\newtheorem{theorem}{Theorem}
\newtheorem{lemma}{Lemma}
\newtheorem*{lemma*}{Lemma}

\newcommand{\norm}[1]{\left\lVert#1\right\rVert}

\makeatletter
\newcommand{\vast}{\bBigg@{3}}
\newcommand{\Vast}{\bBigg@{4}}
\makeatother

\begin{document}
\title{Ergodic Secrecy Rate Analysis\\for LEO Satellite Downlink Networks}

\author{Daeun~Kim,~\IEEEmembership{Student~Member,~IEEE},~and~Namyoon~Lee,~\IEEEmembership{Senior~Member,~IEEE}

\thanks{D. Kim is with the Department of Electrical Engineering, POSTECH, Pohang, Gyeongbuk 37673, South Korea  (e-mail: daeun.kim@postech.ac.kr).}
	\thanks{N. Lee is with the School of Electrical Engineering, Korea University, Seoul 02841, South Korea (e-mail: namyoon@korea.ac.kr).}
}

  \maketitle
\begin{abstract}
Satellite networks are recognized as an effective solution to ensure seamless connectivity worldwide, catering to a diverse range of applications. However, the broad coverage and broadcasting nature of satellite networks also expose them to security challenges. Despite these challenges, there is a lack of analytical understanding addressing the secrecy performance of these networks. This paper presents a secrecy rate analysis for downlink low Earth orbit (LEO) satellite networks by modeling the spatial distribution of satellites, users, and potential eavesdroppers as homogeneous Poisson point processes on concentric spheres. Specifically, we provide an analytical expression for the ergodic secrecy rate of a typical downlink user in terms of the satellite network parameters, fading parameters, and path-loss exponent. Simulation results show the exactness of the provided expressions and we find that optimal satellite altitude increases with eavesdropper density.
\end{abstract}


\section{Introduction}
The low Earth orbit (LEO) satellite networks have undergone remarkable growth and innovation in response to the surging demand for high-speed and ubiquitous connectivity \cite{Zhu2022}. Due to the characteristics of low latency, extensive coverage, and global accessibility, LEO satellite networks are considered indispensable infrastructure for a diverse range of applications. However, the LEO satellite networks are susceptible to potential eavesdropping or jamming due to the wide beam coverage and broadcasting nature. Therefore, the investigation of the secrecy performance of LEO satellite networks is important. Analyzing the secrecy performance in LEO satellite networks provides a comprehensive understanding of the network's security characteristics and practical insights for network operators. 

Stochastic geometry is a useful mathematical tool in characterizing the spatially averaged performance of various wireless networks. By applying Poisson point processes (PPPs) to model the spatial distribution of base stations (BSs) and users, this approach has proven instrumental in gaining valuable insights into the performance of diverse wireless network scenarios \cite{Andrews2011, Lee2015-2}. Especially, performance characterizations considering the physical layer security have been proposed in \cite{Wang2013, Geraci2014, Wang2016}. In \cite{Wang2013}, the achievable secrecy rates are derived by assuming a certain fraction of mobile users as potential eavesdroppers, with a focus on information exchange between BSs and cell association. The secrecy rates achievable through regularized channel inversion precoding are provided for the downlink cellular networks where multiple BSs generate inter-cell interference, and eavesdroppers from neighboring cells collaborate to eavesdrop in \cite{Geraci2014}. For noise-limited millimeter-wave networks, the characterization of secure connectivity probability and the average number of perfect communication links per unit area are presented in \cite{Wang2016}.

In addition to analysis in cellular networks, the PPP-based performance analysis for LEO satellite networks was proposed in \cite{Okati2022, Park2023,park2023_2,lee2022,kim2023}. 

The downlink coverage and rate performance are characterized by modeling satellite locations with a non-homogeneous PPP, where the intensity is determined by the satellites’ actual distribution in \cite{Okati2022}. In \cite{Park2023}, the downlink coverage probability is provided by considering the number of visible satellites on the satellite's spherical cap. In particular, in \cite{Park2023}, the homogeneous PPP-based satellite network model stands out for its mathematical tractability and accuracy. The model not only lends itself to rigorous mathematical analysis but also proves to be a reliable representation of the network, as evidenced by a comparative evaluation of coverage performance with Starlink networks. In \cite{park2023_2}, the modeling and rate coverage probability analysis for satellite-terrestrial integrated networks are proposed. Further, the coverage analysis of satellite networks where satellites are placed on a circular orbit is provided in \cite{lee2022}. The coverage analysis for downlink LEO satellite networks considering the coordinated beamforming is presented in \cite{kim2023}. The aforementioned performance analysis results using PPPs offer insights into various wireless networks. However, none of the above works have successfully analyzed the downlink secrecy performances by considering the satellite's spherical caps, which is the main focus of this paper.

In this paper, we present an analytical expression for the ergodic secrecy rate in downlink LEO satellite networks, where the spatial distribution of satellites, users, and potential eavesdroppers is characterized by a homogeneous spherical Poisson Point Process (SPPP) within concentric spheres. We begin by obtaining a distance distribution between the serving satellite and the nearest eavesdropper from the serving satellite. Leveraging the obtained nearest eavesdropper distance distribution, we derive an exact expression for the ergodic secrecy rate assuming the Shadowed-Riciain fading. The obtained ergodic secrecy rate is expressed in terms of satellite network parameters, path-loss exponent, and channel fading parameters. Through simulation, the accuracy of the derived ergodic secrecy rate is verified. Our finding is that the optimal satellite altitude, which maximizes the ergodic secrecy rate, increases with eavesdropper density. Further, we find that the ergodic secrecy rate varies with satellite density even when the ratio of satellite density to eavesdropper density remains the same. This result is in contrast to the secrecy rate for cellular networks in \cite{Wang2013}, where the secrecy rate remains constant when the ratio of the two densities remains unchanged.

\section{System Model}
This section explains the downlink satellite networks model and the performance metric for analyzing the secrecy performance of the downlink satellite networks.   

\subsection{Satellite Network Model}
\subsubsection{Spatial Distribution of Satellites, Users, and Eavesdroppers} We consider the downlink satellite networks where each satellite sends confidential messages to the users. The satellites are located on the surface of the sphere $\mathcal{S}_{R_{\sf S}}$ having a radius of $R_{\sf S}$, while the downlink users are located on the surface of the Earth $\mathcal{S}_{R_{\sf E}}$ having a radius of $R_{\sf E}$. These satellites are assumed to be spatially distributed according to a homogeneous SPPP $\Phi = \{\mathbf{x}_1,\ldots,\mathbf{x}_N\}$ of density $\lambda$, where $N$ follows Poisson random variable with mean $4\lambda \pi R_{\sf S}^2$. The locations of downlink users are established according to a homogeneous SPPP $\Phi_{\sf u} = \{\mathbf{u}_1,\ldots,\mathbf{u}_M\}$ of density $\lambda_{\sf u}$, where $M$ follows Poisson random variable with mean $4\lambda_{\sf u} \pi R_{\sf E}^2$. We assume that $\Phi_{\sf u}$ is independent of $\Phi$. Further, we consider that the potential eavesdroppers exist on the surface of the Earth $\mathcal{S}_{R_{\sf E}}$. The potential eavesdroppers are spatially distributed according to a homogeneous SPPP $\Phi_{\sf e}=\{\mathbf{e}_1,\ldots,\mathbf{e}_L\}$ of density $\lambda_{\sf e}$, where $L$ follows Poisson random variable with mean $4\lambda_{\sf e}\pi R_{\sf E}^2$. We assume that $\Phi_{\sf e}$ is independent of $\Phi_{\sf u}$.

\begin{figure}[!t]
\centering
\includegraphics[width=6cm]{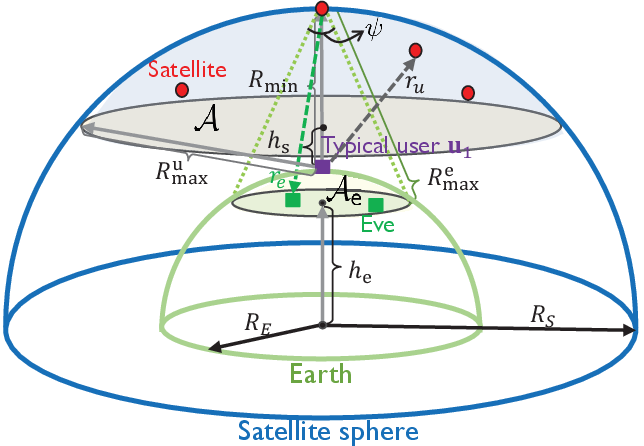}
\caption{An illustration of downlink satellite networks, where the locations of the satellites, users, and potential eavesdroppers are modeled as independent SPPPs.}
\label{fig:network}
\end{figure}

\subsubsection{Visible Spherical Cap of Satellites} By Slivnyak's theorem \cite{Haenggi2005}, we focus on the typical user located at $(0,0,R_{\sf E})$. Then, we define the visible satellite area to the typical user, which is the sphere cap formed by the intersection of the satellite sphere and the tangent plane of an altitude of ${h}_{\sf s}$ above the typical user. We refer to this visible area as a visible spherical cap $\mathcal{A}$, and then the visible satellites within this cap are distributed according to a PPP with density $\lambda |\mathcal{A}|$. The area of the visible spherical cap is computed using Archimedes' hat theorem \cite{cundy1989} as
\begin{align}
    |\mathcal{A}|=2\pi R_{\sf S}(R_{\sf S}-R_{\sf E}-{h}_{\sf s}).
\end{align}

\subsubsection{Spherical Cap of Potential Eavesdroppers} We define the spherical cap $\mathcal{A}_{\sf e}$ where potential eavesdroppers are located within the region of the serving satellite's footprint governed by its beamwidth. This spherical cap is formed by intersecting a plane situated at a distance of ${h}_{\sf e}$ above $(0,0,0)$ to the Earth sphere for $\frac{R_{\sf E}^2}{R_{\sf S}} \le {h}_{\sf e} < R_{\sf E}$. In this case, the distance ${h}_{\sf e}$ corresponds to the satellite's beamwidth $\psi = 2 \arctan\left(\frac{\sqrt{R_{\sf E}^2-h_{\sf e}^2}}{R_{\sf S}-h_{\sf e}}\right)$. Then the potential eavesdroppers within this cap are distributed according to a PPP with density $\lambda_{\sf e}|\mathcal{A}_{\sf e}|$. The area of the visible spherical cap of potential eavesdroppers is computed as
\begin{align}
    |\mathcal{A}_{\sf e}| = 2\pi R_{\sf E}(R_{\sf E}-{h}_{\sf e}).
\end{align}

\subsection{Antenna Gain and Channel Model}
We explain the model of the transmit and receive beamforming gain. We consider the two-lobe approximation of antenna radiation pattern as in \cite{Bai2014}.  In this model, the effective antenna gain is modeled as
\begin{align}
    G = G_{\sf tx}^{\sf main} G_{\sf rx}^{\sf main} \frac{c^2}{(4\pi f_c)^2},~\hat{G} = G_{\sf tx}^{\sf side} G_{\sf rx}^{\sf main} \frac{c^2}{(4\pi f_c)^2},
\end{align}
where $G_{\sf tx}^{\sf main}$ and $G_{\sf rx}^{\sf main}$ are the transmit antenna gain of satellites for the main-lobe and the receive antenna gain of the typical user or eavesdropper for the main-lobe, respectively. Further, $G_{\sf tx}^{\sf side}$ is the transmit antenna gain of satellites for the side-lobe. We assume that the transmit beam of the serving satellite is perfectly oriented to the receive beam of the typical user. Then, $G$ is the effective antenna gain from the serving satellite to the typical user. Furthermore, we consider two scenarios for the transmit antenna gain of the serving satellite concerning the eavesdropper as perfect alignment or misalignment with the eavesdropper's receive beam. Then the effective antenna gain from serving satellite to the eavesdroppers, denoted as $G_{\sf e}$, can be either $G$ or $\hat{G}$.

To model the satellite channels' medium- and small-scale fading effects, we adopt a Shadowed-Rician satellite channel model  \cite{Abdi2003}.
Let $\sqrt{H_{i}^{\sf u}}$ be the fading from $i$th satellite to the typical user and $\sqrt{H_{n}^{\sf e}}$ be the fading from the serving satellite to the $n$th eavesdropper. Then, the probability density function (PDF) of the $\sqrt{H_n^{o}}$ for $o\in\{\sf u, \sf e\}$ is given by
\begin{align}
    f_{\sqrt{H_n^{o}}}(x) &=\left(\frac{2bm}{2bm+\Omega}\right)^m \frac{x}{b}\exp\left(\frac{-x^2}{2b}\right) \nonumber\\ &~~~~~~~~~~~~\cdot F_1\left(m;1;\frac{\Omega x^2}{2b(2bm+\Omega)}\right),
\end{align}
where $F_1(a;b;c)$ is the confluent hyper-geometric function of the first kind \cite{magnus1967}, $m$ is the Nakagami parameter, $2b$ and $\Omega$ are the average power of the scatter component and line-of-sight (LOS) component, respectively.
\subsection{Performance Metric}
We assume that the typical user is served by the nearest satellite and the eavesdropper which is nearest to the serving satellite attempts to eavesdrop. By assuming that there is no interference at downlink receivers, the ergodic secrecy rate for typical user $\mathbf{u}_1$ is given by
\begin{align}
    &R^{\sf sec} =\mathbb{E}\Bigg[\max \Bigg\{\log_2\left(1+\frac{G P H_1^{\sf u} \norm{\mathbf{x}_1-\mathbf{u}_1}^{-\alpha}}{\sigma^2}\right)\nonumber\\& ~~~~~~~ -\log_2\left(1+\frac{G_{\sf e}P H_1^{\sf e} \norm{\mathbf{x}_1-\mathbf{e}_1}^{-\alpha}}{\sigma^2}\right),0\Bigg\}\Bigg],
\end{align}
where $P$ denotes the transmit power of satellites, $\alpha$ denotes a path-loss exponent, and $\sigma^2$ denotes the noise power.  
The complementary cumulative distribution function (CCDF) of $R^{\sf sec}$ conditioned on $\gamma \ge 0$ is given by
\begin{align}
    F^c_{R^{\sf sec}}(\gamma) &= \mathbb{P}\Bigg[\log_2\left(1+\frac{G P H_1^{\sf u} \norm{\mathbf{x}_1-\mathbf{u}_1}^{-\alpha}}{\sigma^2}\right)\nonumber\\&~~~-\log_2\left(1+\frac{G_{\sf e}P H_1^{\sf e} \norm{\mathbf{x}_1-\mathbf{e}_1}^{-\alpha}}{\sigma^2}\right)\ge\gamma \Bigg] \nonumber\\ &=\mathbb{P}\left[\log_2\left(\frac{\Bar{\sigma}^2+H_1^{\sf u} \norm{\mathbf{d}_1}^{-\alpha}}{\Bar{\sigma}^2+\Bar{G}H_1^{\sf e}\norm{\mathbf{d}_e}^{-\alpha}}\right)\ge \gamma\right],
\end{align}
where $\Bar{\sigma}^2=\frac{\sigma^2}{PG}$, $\Bar{G}=\frac{G_{\sf e}}{G}$, $\norm{\mathbf{d}_1}=\norm{\mathbf{x}_1-\mathbf{u}_1}$, and $\norm{\mathbf{d}_e}=\norm{\mathbf{x}_1-\mathbf{e}_1}$. The probability of zero secrecy rate becomes $\mathbb{P}[R^{\sec}=0]=1-F^c_{R^{\sf sec}}(0)$.

\section{Ergodic Secrecy Rate Analysis}
In this section, we provide an exact expression for the ergodic secrecy rate. In the sequel, we use representation of $\tilde{\lambda} = \lambda\frac{R_{\sf S}}{R_{\sf E}}$ and $\Tilde{\lambda}_{\sf e}=\lambda_{\sf e}\frac{R_{\sf E}}{R_{\sf S}}$.

\begin{lemma}[The conditional distribution of nearest eavesdropper distance distribution.]
The conditional distribution of distance $\norm{\mathbf{d}_e}$ between a serving satellite $\mathbf{x}_1$ and the nearest eavesdropper $\mathbf{e}_1$ is given by
\begin{align}
    f_{\norm{\mathbf{d}_e} | \Phi_{\sf e}(\mathcal{A}_{\sf e}) \ge 1} (r_e)=\frac{2\Tilde{\lambda}_e\pi e^{\Tilde{\lambda}_e\pi R_{\sf min}^2}}{1-e^{-\Tilde{\lambda}_e\pi\left(\left(R_{\sf max}^{\sf e}\right)^2-R_{\sf min}^2\right)}}r_e e^{-\Tilde{\lambda}_e\pi r_e^2}, \label{eq:nearest_eve}
\end{align}
for $R_{\sf min} \le r_e\le R_{\sf max}^{\sf e}$, where $R_{\sf min}=R_{\sf S}-R_{\sf E}$ and $R_{\sf max}^{\sf e}=\sqrt{R_{\sf S}^2+R_{\sf E}^2-2R_{\sf S}h_{\sf e}}$.
\begin{proof}
The proof is omitted since it is straightforward from the proof of deriving the nearest satellite distance distribution in \cite{Park2023}.
\end{proof}
\end{lemma}
The derived conditional distribution of the nearest eavesdropper distance is a truncated Rayleigh distribution on the support of $R_{\sf min} \le r_e\le R_{\sf max}^{\sf e}$.

\begin{lemma}[Laplace transform of the eavesdropper's signal power plus noise power.] We define the eavesdropper's signal power plus noise power when $\norm{\mathbf{d}}_e=r_e$ as
\begin{align}
    L_e=\Bar{\sigma}^2+\Bar{G}H_1^{\sf e} r_e^{-\alpha}.
\end{align}
Then, the Laplace transform of $2^{\gamma}L_e -\Bar{\sigma}^2$ under the condition that at least one eavesdropper exists in $\mathcal{A}_{\sf e}$ is given by
\begin{align}
    &\mathcal{L}_{2^{\gamma}L_e -\Bar{\sigma}^2|\Phi_{\sf e}(\mathcal{A}_{\sf e})\ge 1}(s)
    \nonumber \\ & =e^{-s(2^\gamma\Bar{\sigma}^2-\Bar{\sigma}^2)} \sum_{z=0}^{m-1} \zeta(z) \frac{\Gamma(z+1)}{(\beta-c+s 2^\gamma\Bar{G} r_e^{-\alpha})^{z+1}}, \label{eq:laplace}
\end{align}
where $\beta = \frac{1}{2b}$, $c=\frac{\Omega}{2b(2bm+\Omega)}$, $\zeta(z)=\left(\frac{2bm}{2bm+\Omega}\right)^m \frac{\beta (-1)^z (1-m)_z c^z}{(z!)^2}$, $(x)_z=x(x+1)\cdots(x+z-1)$ is the Pochhammer symbol \cite{gradshteyn2014}, and $\Gamma(z)=(z-1)!$ is the Gamma function for positive integer $z$.
\begin{proof}
    The conditional Laplace transform is computed by
    \begin{align}
    &\mathcal{L}_{2^{\gamma}L_e -\Bar{\sigma}^2|\Phi_{\sf e}(\mathcal{A}_{\sf e})\ge 1}(s)
    \nonumber \\&=\mathbb{E}_{H_1^{\sf e}}\left[ e^{-s(2^{\gamma}L_e -\Bar{\sigma}^2)}  \Big|\norm{\mathbf{d}_e}=r_e ,\Phi_{\sf e}(\mathcal{A}_{\sf e}) \ge 1\right]\nonumber \\&=\mathbb{E}_{H_1^{\sf e}}\left[ e^{-s(2^{\gamma}(\Bar{\sigma}^2+\Bar{G}H_1^{\sf e} r_e^{-\alpha})) -\Bar{\sigma}^2)}  \Big|\norm{\mathbf{d}_e}=r_e ,\Phi_{\sf e}(\mathcal{A}_{\sf e}) \ge 1\right] \nonumber\\ &= e^{-s(2^\gamma\Bar{\sigma}^2-\Bar{\sigma}^2)}\mathbb{E}_{H_1^{\sf e}}\!\left[e^{-s 2^\gamma\Bar{G}H_1^{\sf e}r_e^{-\alpha}}\Big| \!\norm{\mathbf{d}_e}\!=\! r_e ,\Phi_{\sf e}(\mathcal{A}_{\sf e}) \ge 1\right].   \label{eq:prf_laplace} 
\end{align}
    By assuming $m$ is integer, the PDF of $H_n^o$ for $o\in\{\sf u,\sf e\}$ is obtained as in \cite{An2016}
    \begin{align}
        f_{H_n^o}(x) = \sum_{z=0}^{m-1} \zeta(z)x^z\exp\left(-(\beta-c)x\right).
    \end{align}
    Then, the inside expectation in \eqref{eq:prf_laplace} is obtained by
    \begin{align}
        &\mathbb{E}_{H_1^{\sf e}}\left[e^{-s 2^\gamma\Bar{G}H_1^{\sf e}r_e^{-\alpha}}\Big|\norm{\mathbf{d}_e}=r_e ,\Phi_{\sf e}(\mathcal{A}_{\sf e}) \ge 1\right]  \nonumber\\ &=\sum_{z=0}^{m-1} \zeta(z) \int_{0}^{\infty}e^{-s 2^\gamma\Bar{G}H_1^{\sf e}r_e^{-\alpha}}\zeta(z)x^z e^{\left(-(\beta-c)x\right)} \mathrm{d}x \nonumber\\ &= \sum_{z=0}^{m-1} \zeta(z) \frac{\Gamma(z+1)}{(\beta-c+s 2^\gamma\Bar{G} r_e^{-\alpha})^{z+1}}. \label{eq:prf_laplace_exp}
    \end{align}
    Plugging \eqref{eq:prf_laplace_exp} into \eqref{eq:prf_laplace}, we obtain the conditional Laplace transform in \eqref{eq:laplace}.
\end{proof}
\end{lemma}

\begin{theorem}
The exact expression for the ergodic secrecy rate is given by
\begin{align}
    &R^{\sf sec} = \int_{0}^{\infty} \vast[\left(1-e^{-\Tilde{\lambda}\pi\left((R_{\sf max}^{\sf u})^2-R_{\sf min}^2\right)}\right)e^{-\Tilde{\lambda}_{\sf e}\pi\left((R_{\sf max}^{\sf e})^2-R_{\sf min}^2\right)} \nonumber\\&\cdot\int_{R_{\sf min}}^{R_{\sf max}^{\sf u}}\Bigg[1-\sum_{z=0}^{m-1} \frac{\zeta(z)\Gamma(z)}{(\beta-c)^{z+1}} \Bigg(1-e^{-(\beta-c)(2^\gamma-1)\bar{\sigma}^2 r_1^{\alpha}}\nonumber\\ &~~~~~\sum_{v=0}^{z}\frac{(\beta-c)^v ((2^\gamma-1)\bar{\sigma}^2 r_1^{\alpha})^v}{v!}\Bigg) \Bigg]f_{\norm{\mathbf{d}_1} | \Phi(\mathcal{A}) \ge 1} (r_1)\mathrm{d}r_1 \nonumber\\&+\left(1-e^{-\Tilde{\lambda}\pi\left((R_{\sf max}^{\sf u})^2-R_{\sf min}^2\right)}\right)\left(1-e^{-\Tilde{\lambda}_{\sf e}\pi\left((R_{\sf max}^{\sf e})^2-R_{\sf min}^2\right)}\right) \nonumber\\ &\cdot\int_{R_{\sf min}}^{R_{\sf max}^{\sf e}}\!\int_{R_{\sf min}}^{R_{\sf max}^{\sf u}} \!\Bigg[1\!-\!\sum_{z=0}^{m-1} \frac{\zeta(z)\Gamma(z)}{(\beta-c)^{z+1}} \Bigg(1 \!-\! \sum_{v=0}^{z}\frac{(\beta\!-\!c)^v (r_1^{\alpha})^v}{v!}\nonumber\\&~~~~ (-1)^v \frac{\mathrm{d}^v \mathcal{L}_{2^{\gamma}L_e -\Bar{\sigma}^2|\Phi_{\sf e}(\mathcal{A}_{\sf e})\ge 1}(s)}{\mathrm{d}s^v}\bigg|_{s=(\beta \!-\!c)r_1^{\alpha}}\Bigg)\Bigg]\nonumber\\ &~~~~~~f_{\norm{\mathbf{d}_1} | \Phi(\mathcal{A}) \ge 1} (r_1)   f_{\norm{\mathbf{d}_e} | \Phi_{\sf e}(\mathcal{A}_{\sf e}) \ge 1} (r_e)   \mathrm{d}r_1\mathrm{d}r_e \vast] \mathrm{d}\gamma.  \label{eq:theorem1}
\end{align}
\begin{proof}
    The CCDF of ergodic secrecy rate is composed by
    \begin{align}
        &F^c_{R^{\sf sec}}(\gamma) \nonumber\\ &= \mathbb{P}[\Phi(\mathcal{A})\ge 1, \Phi_{\sf e}(\mathcal{A}_{\sf e})=0]F^c_{R^{\sf sec}|\Phi(\mathcal{A})\ge 1, \Phi_{\sf e}(\mathcal{A}_{\sf e})=0}(\gamma) \nonumber\\ &+ \mathbb{P}[\Phi(\mathcal{A})\ge 1, \Phi_{\sf e}(\mathcal{A}_{\sf e})\ge1]F^c_{R^{\sf sec}|\Phi(\mathcal{A})\ge 1, \Phi_{\sf e}(\mathcal{A}_{\sf e})\ge 1}(\gamma) \nonumber\\ &\overset{(a)}{=} \mathbb{P}[\Phi(\mathcal{A})\ge 1]\mathbb{P}[\Phi_{\sf e}(\mathcal{A}_{\sf e})=0]F^c_{R^{\sf sec}|\Phi(\mathcal{A})\ge 1, \Phi_{\sf e}(\mathcal{A}_{\sf e})=0}(\gamma) \nonumber\\ &+ \mathbb{P}[\Phi(\mathcal{A})\ge 1]\mathbb{P}[\Phi_{\sf e}(\mathcal{A}_{\sf e})\ge1]F^c_{R^{\sf sec}|\Phi(\mathcal{A})\ge 1, \Phi_{\sf e}(\mathcal{A}_{\sf e})\ge 1}(\gamma),   \label{eq:ccdf_composition}
    \end{align}
    where (a) follows from the independence of $\Phi$ and $\Phi_{\sf e}$. Then, the ergodic secrecy rate is obtained by integrating \eqref{eq:ccdf_composition} from 0 to $\infty$ with respect to $\gamma$ as follows:
    \begin{align}
        &R^{\sf sec} = \int_{0}^{\infty}F^c_{R^{\sf sec}}(\gamma)\mathrm{d}\gamma \nonumber\\ &= \int_{0}^{\infty}\!\! \Big(\mathbb{P}[\Phi(\mathcal{A})\ge 1]\mathbb{P}[\Phi_{\sf e}(\mathcal{A}_{\sf e})=0]F^c_{R^{\sf sec}|\Phi(\mathcal{A})\ge 1, \Phi_{\sf e}(\mathcal{A}_{\sf e})=0}(\gamma) \nonumber\\ &+ \mathbb{P}[\Phi(\mathcal{A})\ge 1]\mathbb{P}[\Phi_{\sf e}(\mathcal{A}_{\sf e})\ge1]F^c_{R^{\sf sec}|\Phi(\mathcal{A})\ge 1, \Phi_{\sf e}(\mathcal{A}_{\sf e})\ge 1}(\gamma)\Big)\mathrm{d}\gamma. \label{eq:gamma_int}
    \end{align}
    
    By assuming $m$ is integer, the CDF of $H_n^o$ for $o\in\{\sf u,\sf e\}$ is obtained as
    \begin{align}
        &F_{H_n^o}(x) = \sum_{z=0}^{m-1} \frac{\zeta(z)\Gamma(z)}{(\beta-c)^{z+1}} \left(1\!-\!e^{-(\beta-c)x}\sum_{v=0}^{z}\frac{(\beta-c)^v x^v}{v!}\right).
    \end{align}

When $\Phi(\mathcal{A})\ge 1$ and $\Phi_{\sf e}(\mathcal{A}_{\sf e})=0$, the ergodic secrecy rate becomes the ergodic rate of the typical user. Then, the CCDF of the ergodic secrecy rate is derived as
\begin{align}
    &F^c_{R^{\sf sec}|\Phi(\mathcal{A})\ge 1, \Phi_{\sf e}(\mathcal{A}_{\sf e})=0}(\gamma)\nonumber\\&=\mathbb{E}_{r_1}\bigg[\mathbb{P}\bigg[\log_2\left(1+\frac{H_1^{\sf u} r_1^{-\alpha}}{\bar{\sigma}^2}\right)\ge \gamma \nonumber\\ &~~~~~~~~~~~~~~~~~~~~~~~~ \bigg| \Phi(\mathcal{A})\ge 1, \Phi_{\sf e}(\mathcal{A}_{\sf e})= 0, \norm{\mathbf{d}_1}=r_1\bigg]\bigg] \nonumber\\&= \mathbb{E}_{r_1}\Bigg[1-\sum_{z=0}^{m-1} \frac{\zeta(z)\Gamma(z)}{(\beta\!-\!c)^{z+1}} \Bigg(1-e^{-(\beta-c)(2^\gamma-1)\bar{\sigma}^2 r_1^{\alpha}}\nonumber\\ &~~~~~\sum_{v=0}^{z}\frac{(\beta\!-\!c)^v ((2^\gamma-1)\bar{\sigma}^2 r_1^{\alpha})^v}{v!}\Bigg)  
    \Bigg| \Phi(\mathcal{A})\ge 1, \Phi_{\sf e}(\mathcal{A}_{\sf e})= 0\Bigg] \nonumber\\
    &=\int_{R_{\sf min}}^{R_{\sf max}^{\sf u}}\Bigg[1-\sum_{z=0}^{m-1} \frac{\zeta(z)\Gamma(z)}{(\beta-c)^{z+1}} \Bigg(1-e^{-(\beta-c)(2^\gamma-1)\bar{\sigma}^2 r_1^{\alpha}}\nonumber\\ &~~~~~\sum_{v=0}^{z}\frac{(\beta-c)^v ((2^\gamma-1)\bar{\sigma}^2 r_1^{\alpha})^v}{v!}\Bigg) \Bigg]f_{\norm{\mathbf{d}_1} | \Phi(\mathcal{A}) \ge 1} (r_1)\mathrm{d}r_1, \label{eq:pf_ccdf1}
\end{align}
where the conditional nearest satellite distance distribution $f_{\norm{\mathbf{d}_1} | \Phi(\mathcal{A}) \ge 1} (r_1)$ has been derived in \cite{Park2023} as
\begin{align}
    f_{\norm{\mathbf{d}_1} | \Phi(\mathcal{A}) \ge 1} (r_1)=\frac{2\Tilde{\lambda}\pi e^{\Tilde{\lambda}\pi R_{\sf min}^2}}{1-e^{-\Tilde{\lambda}\pi((R_{\sf max}^{\sf u})^2-R_{\sf min}^2)}}r_1 e^{-\Tilde{\lambda}\pi r_1^2},
\end{align}
for $R_{\sf min} \le r_1\le R_{\sf max}^{\sf u}$, where $R_{\sf max}^{\sf u}=\sqrt{R_{\sf S}^2-R_{\sf E}^2-2R_{\sf E}h_{\sf e}}$.

Further, the CCDF of the ergodic secrecy rate when $\Phi(\mathcal{A})\ge 1$ and $\Phi_{\sf e}(\mathcal{A}_{\sf e})\ge 1$ is given by
\begin{align}
    &F^c_{R^{\sf sec}|\Phi(\mathcal{A})\ge 1, \Phi_{\sf e}(\mathcal{A}_{\sf e})\ge 1}(\gamma)\nonumber \\&=\mathbb{E}_{r_1,r_e}\bigg[\mathbb{P}\bigg[\log_2\left(\frac{\Bar{\sigma}^2+H_1^{\sf u} r_1^{-\alpha}}{L_e}\right)\ge \gamma \nonumber\\ &~~~~~~~ \bigg| \Phi(\mathcal{A})\ge 1, \Phi_{\sf e}(\mathcal{A}_{\sf e})\ge 1, \norm{\mathbf{d}_1}=r_1,\norm{\mathbf{d}_e}=r_e\bigg]\bigg] \nonumber\\ &=\mathbb{E}_{r_1,r_e}\Bigg[\mathbb{E}_{H_1^{\sf e}}\Bigg[1\!-\!\sum_{z=0}^{m-1} \frac{\zeta(z)\Gamma(z)}{(\beta-c)^{z+1}} \Bigg(1\!-\! \sum_{v=0}^{z}\frac{(\beta-c)^v (r_1^{\alpha})^v}{v!} \nonumber\\&(2^{\gamma}L_e -\Bar{\sigma}^2 )^v e^{-(\beta-c)r_1^{\alpha}(2^{\gamma}L_e -\Bar{\sigma}^2)}\Bigg)\bigg|\Phi(\mathcal{A})\ge 1, \Phi_{\sf e}(\mathcal{A}_{\sf e})\ge 1,\nonumber\\& \norm{\mathbf{d}_1}=r_1,\norm{\mathbf{d}_e}=r_e \Bigg]\Bigg|\Phi(\mathcal{A})\ge 1, \Phi_{\sf e}(\mathcal{A}_{\sf e})\ge 1\Bigg] \nonumber\\ &\overset{(a)}{=}\mathbb{E}_{r_1,r_e}\Bigg[1\!-\!\sum_{z=0}^{m-1} \frac{\zeta(z)\Gamma(z)}{(\beta-c)^{z+1}} \Bigg(1\!-\! \sum_{v=0}^{z}\frac{(\beta-c)^v (r_1^{\alpha})^v}{v!}(-1)^v\nonumber\\&  \frac{\mathrm{d}^v \mathcal{L}_{2^{\gamma}L_e -\Bar{\sigma}^2|\Phi_{\sf e}(\mathcal{A}_{\sf e})\ge 1}(s)}{\mathrm{d}s^v}\bigg|_{s=(\beta\!-\!c)r_1^{\alpha}}\Bigg) \Bigg|\Phi(\mathcal{A})\!\ge\! 1, \Phi_{\sf e}(\mathcal{A}_{\sf e})\!\ge\! 1\Bigg] \nonumber\\ &\overset{(b)}{=}\int_{R_{\sf min}}^{R_{\sf max}^{\sf e}}\!\int_{R_{\sf min}}^{R_{\sf max}^{\sf u}} \!\Bigg[1\!-\!\sum_{z=0}^{m-1} \frac{\zeta(z)\Gamma(z)}{(\beta-c)^{z+1}} \Bigg(1 \!-\! \sum_{v=0}^{z}\frac{(\beta\!-\!c)^v (r_1^{\alpha})^v}{v!}\nonumber\\&~~~~ (-1)^v \frac{\mathrm{d}^v \mathcal{L}_{2^{\gamma}L_e -\Bar{\sigma}^2|\Phi_{\sf e}(\mathcal{A}_{\sf e})\ge 1}(s)}{\mathrm{d}s^v}\bigg|_{s=(\beta \!-\!c)r_1^{\alpha}}\Bigg)\Bigg]\nonumber\\ &~~~~~~~~~f_{\norm{\mathbf{d}_1} | \Phi(\mathcal{A}) \ge 1} (r_1)   f_{\norm{\mathbf{d}_e} | \Phi_{\sf e}(\mathcal{A}_{\sf e}) \ge 1} (r_e)  \mathrm{d}r_1\mathrm{d}r_e, \label{eq:pf_ccdf2}
\end{align}
where (a) follows from applying the derivative property of the Laplace transform, i.e., $\mathbb{E}\left[X^v e^{-sX}\right]=(-1)^v \frac{\mathrm{d}\mathcal{L}_{X}(s)}{\mathrm{d}s^v}$ and (b) follows from the independence of distance distribution $f_{\norm{\mathbf{d}_1} | \Phi(\mathcal{A}) \ge 1} (r_1)$ and $f_{\norm{\mathbf{d}_e} | \Phi_{\sf e}(\mathcal{A}_{\sf e}) \ge 1} (r_e)$. 

By plugging \eqref{eq:pf_ccdf1} and \eqref{eq:pf_ccdf2} into \eqref{eq:gamma_int}, we finally obtain the ergodic secrecy rate in \eqref{eq:theorem1}, which completes the proof.

\end{proof}
\end{theorem}

\section{Simulation Results}
In this section, we provide simulation results on the ergodic secrecy rate to validate the derived expressions and evaluate the performance according to the network parameters. We set $\alpha=2$, $[m,b,\Omega]=[1,0.063,8.97 \times 10^{-4}]$, $P=43$ dBm, $N_0=-174$ dbm/Hz, $W=100$ MHz, $G_{\sf tx}^{\sf main}=G_{\sf rx}^{\sf main}=30$ dBi, $G_{\sf tx}^{\sf side}=20$ dBi, $h_{\sf e}=\frac{R_{\sf E}^2}{R_{\sf S}}$, and $h_{\sf s}=0$ km.

\begin{figure}[!t]
    \centering
    \includegraphics[width=7cm]{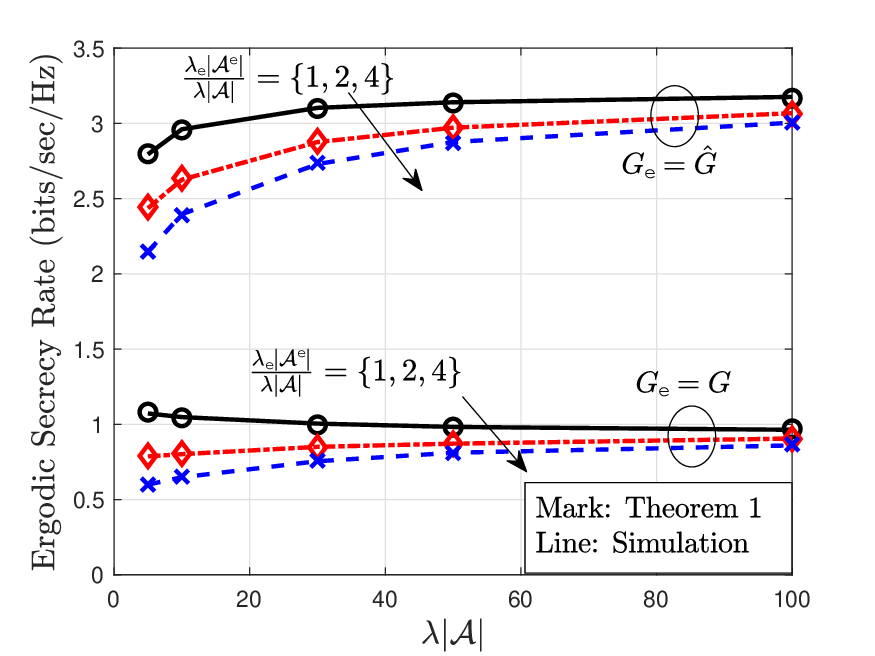}
    \caption{The ergodic secrecy rate $R^{\sf sec}$ according to the ratio of satellite density to eavesdropper density.} \label{fig:ratio}
\end{figure}

\begin{figure}[t]
    \centering
    \includegraphics[width=7cm]{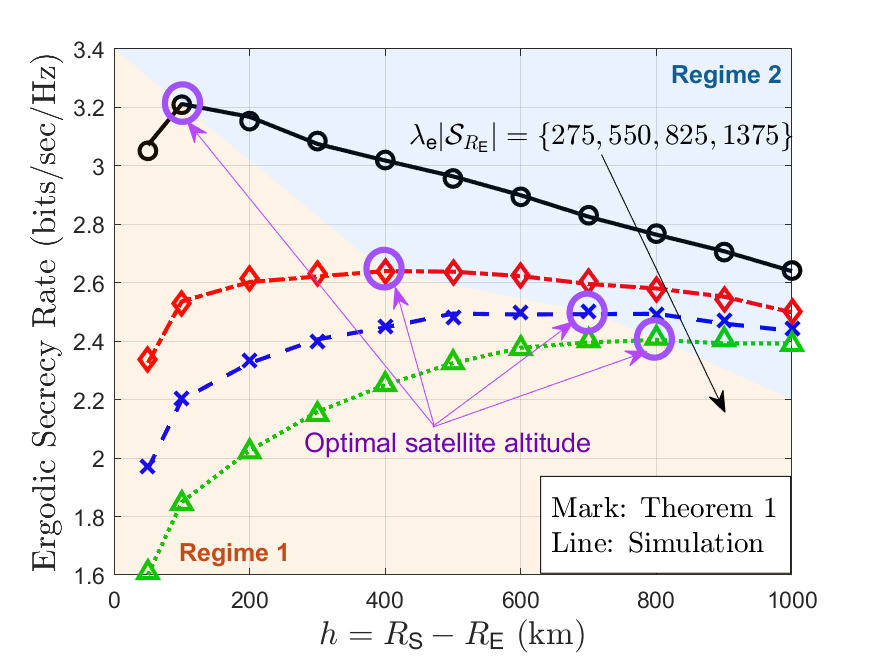}

    \caption{The ergodic secrecy rate $R^{\sf sec}$ versus satellite altitude $h$ for various eavesdropper densities.} \label{fig:altitude}
\end{figure}


Fig. \ref{fig:ratio} shows the ergodic secrecy rate according to the ratio of satellite density and eavesdropper density for $h=R_{\sf S}-R_{\sf E}=500$ km. We consider two scenarios where the eavesdropper is located within the main-lobe beam of the serving satellite $(G_{\sf e}=G)$, and the eavesdropper is located within the side-lobe beam of the serving satellite $(G_{\sf e}=\hat{G})$. As shown in Fig. \ref{fig:ratio}, the derived ergodic secrecy rate in Theorem 1 exactly matches the numerical ergodic secrecy rate.  The ergodic secrecy rate is lower when the eavesdropper is located in the main-lobe beam region compared to that in the side-lobe beam region of the serving satellite.  Further, when increasing the eavesdropper density, the ergodic secrecy rate decreases. However, even when the ratio of satellite density to eavesdropper density is equivalent, the ergodic secrecy rate varies according to the density itself. This result contrasts with the behavior of the secrecy rate in terrestrial cellular networks, where the secrecy rate depends solely on the ratio of BS density to eavesdropper density. The distinction arises from our satellite network model, which considers the spatial distribution of satellite and eavesdropper locations in finite space, differing from the analysis in cellular networks that assume an infinite two-dimensional space.

Fig. \ref{fig:altitude} shows the ergodic secrecy rate according to the satellite altitude $h$. In this simulation, we fix the total number of satellites that can be deployed on $\mathcal{S}_{R_{\sf S}}$ as 275, which corresponds to $\lambda |\mathcal{A}|=10$ for $h=500$ km. As depicted in Fig. \ref{fig:altitude}, the optimal satellite altitude increases with the density of eavesdroppers. When the density of eavesdroppers is low, it becomes crucial to increase the SNR of the desired link by reducing the satellite altitude. Conversely, when the eavesdropper density is high, it is more advantageous to diminish the SNR of the eavesdroppers by elevating the satellite altitude. Nevertheless, excessive increases in satellite altitude result in a significantly reduced SNR for the desired link, leading to a lower ergodic secrecy rate as can be seen in regime 2. Therefore, the choice of satellite altitude must be made judiciously to maximize the secrecy rate.

\section{Conclusion}
We have introduced an analytical expression for the ergodic secrecy rate in downlink LEO satellite networks. Using stochastic geometry tools, we derived an exact expression for the ergodic secrecy rate assuming Shadowed-Riciain fading. The derived ergodic secrecy rate is expressed in terms of satellite network parameters, path-loss exponent, and channel fading parameters. 
Our key finding is that the optimal satellite altitude exhibits an increasing trend with higher eavesdropper density. Additionally, we find that the secrecy rate varies with satellite density even when the density ratio remains constant. This contrasts with cellular networks where the secrecy rate remains constant despite changes in the density ratio. These insights provide valuable implications for optimizing satellite network configurations in the presence of varying eavesdroppers and satellite densities.

\bibliographystyle{IEEEtran}
\bibliography{IEEEabrv,Reference}

\end{document}